# Nonlinear Optics: a look from the interaction time viewpoint and what it portends.

Jacob B Khurgin

Johns Hopkins University, Baltimore MD 21218 USA

I present a simple view of nonlinear optical phenomena as being determined mostly by the length of interaction time between photons and matter. This may explain why in the last decades the progress in developing better nonlinear materials has been not as rapid as wished. A few tentative routes towards possible improvements in the efficiency of nonlinear optical phenomena are suggested.

## Introduction

The field of Nonlinear Optics as we know it, was born shortly after the first demonstrations of lasers in early 1960's [1], and the first treatise on nonlinear optics, "*Nonlinear Optics*" by Bloembergen appeared in 1965 [2], followed, two decades later by a more comprehensive treatment by Shen, "*Principles of Nonlinear Optics*"[3]. By that time the fundamental science of Nonlinear Optics had been developed and most of the second and third order nonlinear phenomena well understood and experimentally characterized. If one takes a look at the first 1984 edition of this book, one can see that the ensuing 40 years has not added much to the information already contained in that book, i.e., forty years on, one still keeps reading scientific papers about revolutionary applications such as all optical switching[4], and ultrafast signal processing [5] in endless flow of new exotic materials, while for the relatively limited list of practical applications (harmonic , optical parametric, continuum, and optical frequency comb generation pretty much exhausts this list) one still relies on a handful of well-developed nonlinear crystals (LiNbO$_3$[6], KTP[7], BBO[7], metal chalcogenides [8]) and on optical fibers [9, 10] and Si waveguides [11]. The situation is strikingly different from many other areas of photonics say lasers [12, 13] where the rapid progress has made many otherwise excellent 40-year-old books limited in scope as they would contain no mentioning of the devices that are mainstay of photonics today , such as quantum cascade, Yb-fiber, Ti-Sapphire, or GaN blue lasers and VCSELs to name a few). Past years have seen ebb and flow of news about new promising nonlinear materials or schemes. Just to mention a few:

- nonlinear polymers [14],
- semiconductor quantum wells[15],
- quantum dots[16],
- nanotubes [17],
- nonlinear photonic crystals[18],
- nonlinear plasmonics [19],
- slow light[20, 21],
- electro-magnetically induced transparency[22],
- nonlinear metamaterials (both metallic and dielectric)[23],
- metasurfaces[24],
- graphene[25, 26],
- other two-dimensional materials such as transition metal dichalcogenides [27]
- perovskites [28, 29]
- Weyl semi-metals [30]
- epsilon-near-zero materials[31] ,

- topological photonics[32],

and the list may go on. With a few notable exceptions (photonic crystal [33] and tapered fibers [34] for continuum generation and various new materials for saturable absorbers) the research on this set of topics, while greatly extending our horizons, has yielded relatively modest results in the practical sense, and the long foretold all optical digital computer [35] has not materialized, while the jury is still out on all-optical neural networks also requiring high nonlinearity[36]. If anything, the among the greatest practical successes of nonlinear optics over the last two decades have been advances in the areas where nonlinear effects are deleterious, e.g., in development of techniques to mitigate nonlinearities in high-capacity fiber optical communication links [37] or the means to avoid self-focusing and filamentation of powerful laser beams.[38] (I am not touching on completely different and very successful field of nonlinear spectroscopy [39] here). With six decades on nonlinear optics behind us, in my view, it would be worthwhile to take a fresh look at the limits of nonlinear optics and try to develop a unified picture of the diverse nonlinear phenomena, with a goal of providing engineers with intuitive understanding of what may and what may not possible.

Before continuing, I want to emphasize that this discourse is neither a comprehensive review nor a tutorial. Also, it is not an attempt to divine the future of all-optical processing and other mind-catching applications as, sadly, I do not belong to the category of "visionary" scholars who can foresee what future portends. While the visionary scientist's role is alike that of an architect conceiving inspiring designs of magnificent edifices, I see my role as that of a structural engineer coldly assessing whether the conceived design will stand on its own and the roof will not leak at first rain. So, what follows is nothing but my own, and very subjective at that, opinion formed by more than four decades of experience, opinion with which many readers may disagree. Some of the analysis presented here is known, so a better-informed reader may skip it, but a significant part is original and may be useful to the reader, a novice, or an old hand alike. Also, it is worth mentioning that here I focus exclusively on nonlinearities associated with the **real** part of the refractive index (i.e., switching, modulation and frequency conversion), because it is these phenomena that tend to arouse imagination and portend all-optical networks in which the optical signals shuffle smoothly between the optical components while changing wavelengths and modulation formats with lightning speed, and performing these feats with not a single wire attached. Thus, important, even though less futuristic and more developed areas of saturable absorption[40] for mode-locking, nonlinear[39] and Raman[41] spectroscopies, are left out of the scope. Also left out of scope is the nonlinearities in the THz region[42] because this region is closer to the electronic rather than optical domain, and electronic devices operating in this region, such as Schottky diodes[43] and high electron mobility transistors [44] have nonlinear response that dwarfs that of any conceivable nonlinear material.

## What is a proper figure of merit for nonlinearity (or why not divide by zero?)?

The established metric for the nonlinear properties is nonlinear susceptibility $\chi^{(n)}$, and in case of third order nonlinearity nonlinear index $n_2$ closely related to $\chi^{(3)}$. Essentially, either one of these measures refers to the permittivity change imposed by unit electric field ($\chi^{(2)}E$) or square of it ($\chi^{(3)}E^2$). If modulation of index is time dependent, this modulation results in frequency conversion. However, the index modulation per se means very little unless it is accomplished over sufficient length. Indeed, the efficiency of frequency conversion increases as $\sin^2(\chi^{(n)}E^{n-1}k_0L/2n)$ where $k_0 = \omega/c$ [3]. Nonlinear index change is $\Delta n = \Delta\varepsilon/2n = \chi^{(n)}E^{n-1}/2n$, so conversion efficiency changes as $\sin^2(\Delta\Phi_{nl})$ where nonlinear phase shift is $\Delta\Phi_{nl} = \Delta n k_0 L$. Thus achieving 100% conversion efficiency in sum or difference frequency generation, or in four-wave mixing, just as achieving 100% switching in nonlinear interferometer or coupler requires nonlinear phase shift on the order of at least $\pi$ (or $\pi/2$ for push-pull arrangement) as shown schematically in Fig.1a

While this fact is well known, perhaps it is less recognized that this relation follows from the uncertainty principle and thus cannot be violated. Indeed, change of index is tantamount to change of momentum, $\Delta p = \Delta n \hbar k_0$, and in order to measure, or register this change, the interaction length $L$ over which the measurement is performed is such that the product $\Delta p L \geq \hbar/2$, which immediately postulates $\Delta\Phi_{nl} \geq 1/2$ as shown in Fig.1b. The equality sign corresponds to Gaussian probability distribution and for uniform distribution relating to propagation with constant velocity $\Delta\Phi_{nl} \sim \pi/2$, appears a reasonable condition (from the start, I want to state that all the analysis to be presented here is just an order of magnitude and the goal is not to provide a "design guide" but simply to outline the boundaries of the realm of the possible)

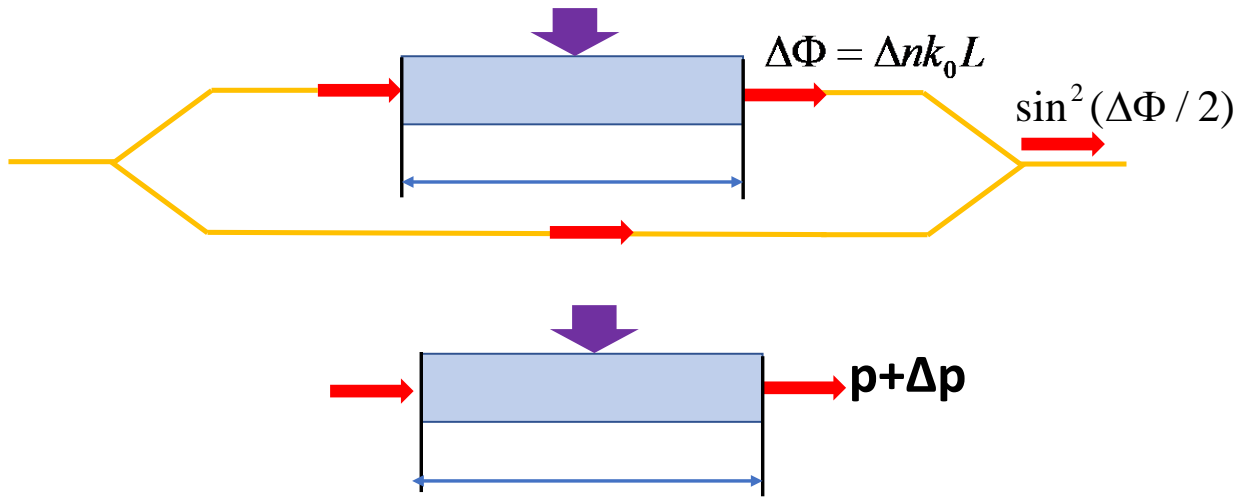

**Figure 1. (a)** An example of arrangement for nonlinear switching of signal light by a control light using Mach Zehnder Interferometer. One can use all types of interferometers or filters with essentially the same requirement for efficient switching $\Delta\Phi_{nl} \sim \pi/2$. **(b)** Converting nonlinear phase shift into intensity switching is equivalent to measuring the momentum change $\Delta p = \Delta n \hbar k_0$ and requires minimum interaction length in accordance with Heisenberg's uncertainty principle.

Therefore, from the practical point of view the value of nonlinear susceptibility $\chi^{(n)}$ is far less relevant than maximum nonlinear phase shift attainable for a given intensity of light $\Delta\Phi_{nl}$, or, alternatively the intensity required to get 90 degrees nonlinear phase shift $I_{\pi/2}$. Obviously, the phase shift is limited by the interaction length, which, for various nonlinearities is limited by absorption length, by fabrication, or simply by the desire to limit the size of individual device.

This fact puts into proper perspective the numerous reports of "giant" nonlinearities in two-dimensional materials [45, 46], metal films[47], and various metasurfaces[48]. Taking vanishingly small nonlinear output and dividing it by very small thickness (a few angstroms in case of 2D materials) may indeed yield values of $\chi^{(n)}$ that warrant press releases and high impact publications, but these claims cannot overturn the stubborn fact that the efficiency is still vanishingly small, and increasing interaction length to boost the efficiency is far from trivial. For instance, going from monolayer to thicker structures often changes the entire character of the material, e.g., direct bandgap becomes indirect and so on.

## Intuitive understanding of why nonlinear phenomena are so weak.

Having settled the issue of boosting nonlinear susceptibility using the magic of division by zero, let us turn attention to the issue of why the nonlinearity is, well, so weak? On the most fundamental level the answer is simple: photons happen to be bosons that carry no charge and are not subject to Pauli exclusion principle, so they do not interact with each other directly, but only via intermediary short-lived states of the matter (typically electrons and holes). One can also frame this answer in a different way, that that photons in the medium are no longer pure photons, but polaritons – coupled field-matter excitations (quasi-particles) that do interact with each other albeit very weakly. The fact that the interactions are so weak is the main reason why photons are just about perfect for unimpeded long-distance transmission of information, but, at the same time it makes manipulation of information with photons a daunting task, in confirmation of the universal principle of unavailability of free lunch.

One way to quantify the weakness of nonlinear interactions is to consider a typical anharmonic potential as shown in Fig.2a which may describe binding potential $U(r)$ of an electron in a chemical bond of characteristic size $a$ (of the order of 1-3A for bonds active in optical range) with binding energy $U_0$ (on the scale of a few eV). For small deviations from the equilibrium position the potential is very well described by harmonic (i.e., parabolic) function shown as a dashed line. One can loosely define this region as "linear". At the same time, with displacement approaching $a/2$ electron is no longer strongly bound, and one finds the electron in the "optical breakdown" region. As one can see, the" nonlinear" region is squeezed between the linear and breakdown regions and it only when one is on the verge of breakdown that nonlinear change of permittivity becomes commensurate with the linear one.

In this most primitive model, polarizability as well as the linear susceptibility $\chi^{(1)} = \varepsilon - 1$ depend on the curvature of the potential and the curvature changes substantially (factor of ~2) when the energy of electron is commensurate with $U_0$. To achieve roughly 100% change in susceptibility either each valence electron must acquire energy commensurate with $U_0$, or, one can also change curvature by factor of ~2 by distorting the lattice so that $U_0$ is changed by that factor. Either way, one can then make a following conjecture – if the energy density stored (not necessarily absorbed!) in the material is roughly $\Delta u_{st} \sim u_0 = NU_0$, where $N$ is the density of valence electrons, then the electronic part of susceptibility $\chi^{(1)} = \varepsilon - 1$ changes by roughly 100% and it follows that change of index can be roughly estimated as $\Delta n / n \sim \frac{1}{2} \Delta \varepsilon / \varepsilon \sim \Delta u_{st} / u_0$.

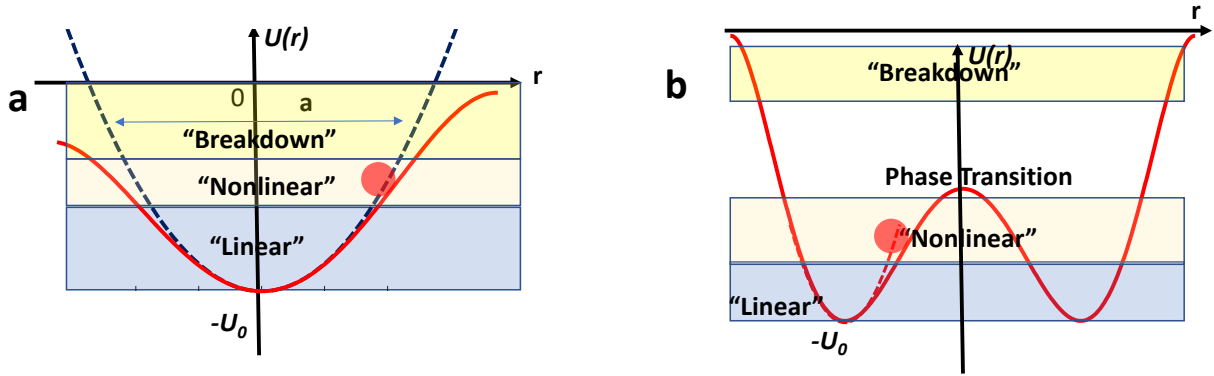

**Figure 2**. **(a)** Anharmonic potential showing that nonlinearity becomes appreciable only close to optical breakdown. **(b)** Anharmonic potential in the vicinity of phase transition

The question of what value of energy should be taken as $U_0$ does not have an unambiguous answer, but it should be noted that in the most simple potential well model the depth $U_0$ is commensurate with the split between the lowest and the first excited statse – in semiconductors and dielectrics this split corresponds to the "mean" splitting of valence and conduction band and commonly referred to as Penn gap[49], which is always in the range of a few eV for the materials transparent in optical/ near IR range. On the higher end of the estimate, one can consider cohesive energy (per atom) [50]as $U_0$ - and that value is typically closer to 5-10 eV, while when it comes to the nonlinearities associated with nonparabolicities in the bands[51], it is roughly the fundamental direct bandgap energy that should be considered, which is somewhat lower than Penn gap. However, these energies (as well as electron affinity and ionization energy) are all on the same 1-10eV scale, and, furthermore, for the materials transparent over the wider range of frequencies (typically having more ionic character of bonding) all these energies tend to increase. Therefore, the order-of-magnitude assessments performed here are valid no matter what is chosen as $U_0$ Since the density of valence electrons is $N \sim 1-2 \times 10^{23} cm^{-3}$ and $U_0 \sim 1-10 eV$ it seems that the energy density required to change refractive index by 100% is on the scale of $u_0 \sim 10^4 - 3 \times 10^5 J/cm^3$.

To verify the sanity of this analysis, consider *very diverse* mechanisms of changing refractive index using thermal, electro-optic, acousto-optic, $\chi^{(2)}$ and $\chi^{(3)}$. The examples are shown in Table I, and despite the mechanisms being so different, the range $u_0$ corresponds to the one obtained above from quite general intuitive considerations.

| Modulation Method | Index change. Energy density | Expression for $u_0$ | Material | Material parameters | Value of $u_0$ (J/cm³) |
|---|---|---|---|---|---|
| **Thermal** | $\Delta n = (dn/dT)\Delta T$ <br> $\Delta u_{st} = c_v \rho \Delta T$ | $u_0 = nc_v\rho/(dn/dT)$ | Si[52] | $n = 3.47$, <br> $\rho = 2.3 g/cm^3$ <br> $c_V = 0.7 J/g \cdot K$ <br> $dn/dT = 1.8 \times 10^{-4} K^{-1}$ | $3.1 \times 10^4$ |
| **Electro-** | $\Delta n = n^3 r_{33} E_{DC}/2$ | $u_0 = 2\varepsilon_0 \varepsilon / n^4 r_{33}^2$ | LiNbO$_3$[53] | $n = 2.2$, | $2.4 \times 10^4$ |
| optical | $\Delta u_{st} = \varepsilon_0 \varepsilon E_{DC}^2 / 2$ | | | $r_{33} = 30\,pm/V$<br>$\varepsilon = 28.7$ | |
| Optical Kerr | $\Delta n = n_2 I$<br>$\Delta u_{st} = In/c$ | $u_0 = n^2/cn_2$ | Si[54] | $n_2 = 3.4 \times 10^{-14}\,cm^2/W$ | $1.2 \times 10^4$ |
| 2$^{nd}$ order nonlinear | $\Delta n = \chi^{(2)} E_\omega / 2n$<br>$\Delta u_{st} = \varepsilon_0 n^2 E_\omega^2 / 2$ | $u_0 = \varepsilon_0 n^6/2\left|\chi^{(2)}\right|^2$ | GaAs[55] | $n = 3.4$,<br>$\chi^{(2)} \sim 200\,pm/V$ | $18.1 \times 10^4$ |
| Acousto-optic | $\Delta n = n^3 PS/2$<br>$\Delta u_{st} = \rho v_s^2 S^2/2$ | $u_0 = 2n^2/M_2 v_s$ | TeO$_2$ [56] | $n = 2.1$<br>$M_2 = 35 \times 10^{-15}\,s^3/kg$<br>$v_s = 4,250\,m/s$ | $6.9 \times 10^4$ |

**Table 1.** Examples of energy density $u_0$ required to achieve ~100% change of refractive index in the optical region using various modulation methods. $n$-refractive index, $dn/dT$ thermal optic coefficient, $c_V$-specific heat, $\rho$-mass density, $r_{33}$-Pockels coefficient, $E_{DC}$ – low frequency electric field, $\varepsilon$-static dielectric constant, $n_2$ – nonlinear refractive index, $I$-optical power density, $c$-speed of light, $\chi^{(2)}$ second order susceptibility, $E_\omega$-optical field, $S$-strain, $P$-elasto-optic coefficient, $M_2 = n^6 P^2/\rho v_s^3$-acousto-optic figure of merit, $v_s$ sound velocity.

Let us now think for a moment what is the meaning of this remarkable results – the *energy density* required to change index by a given amount $\Delta n$ is $\Delta u_{st} \sim (\Delta n/n) u_0$ which is more or less similar in every nonlinear material (for the same wavelength range), but the *power density* $I = \Delta u_{st}/\tau$ can be changed by orders of magnitude by changing the "energy storage" time, $\tau$. This time can be the lifetime of electron on a given energy level or band, thermal diffusion time, photon lifetime in the resonator, or photon propagation time among many others. As we shall see, *engineering the storage or interaction time $\tau$ and accepting (almost inevitable) bandwidth limitation is what all the conceivable pathways to higher nonlinearity amount to in the end.*

Before continuing with impact of interaction time, let us estimate the scale of nonlinear susceptibility the same model of Fig.1a. If one defines the binding ("atomic") field $E_0 \sim U_0/ea$ [57], it follows then that when applied optical field approaches $E_0$, the displacement becomes comparable to $a$ and nonlinear change of permittivity is on the scale of 100%. Therefore, the scale of nonlinear susceptibility is $\chi^{(n)} \sim (\varepsilon - 1)/E_0^{n-1}$. For a typical solid state medium linear susceptibility $\chi^{(1)} = \varepsilon - 1$ is on the order of 2-20 with lower values and corresponding to materials transparent all the way into UV and higher one to the materials transparent only in the IR range. Note that this expression assumes that the entire lattice participates in optical process – if one is looking for the response of a material due to dopant then nonlinear susceptibility gets scaled down by the fraction of "active" entities.

Since the order of magnitude of $E_0$ is between $10^{10}$ and $10^{11}$ V/m and permittivity ranges between 2 and 20, typical value of $\chi^{(3)}$ is $10^{-18} - 10^{-22}\,m^2/V^2$ corresponding to the nonlinear index $n_2 \sim 10^{-13} - 10^{-16}\,cm^2/W$. Note that typically material with wide bandgap, transparent throughout visible and near UV range are stronger bound, i.e., the energy $U_0$ in them is higher and the bond length is shorter, hence the nonlinear index in them is lower than in narrower bandgap materials, that are transparent only in IR region. *Therefore, when comparing different nonlinear materials, it is important to compare them in the same spectral region.* When doing that, one, non-surprisingly discovers that for materials with comparable transparency ranges the off-resonant (i.e., not associated with absorption) nonlinear indices are also comparable with not much room left for engineering "giant" nonlinearities.

For the second (and other even) order nonlinearity the situation is more complex as lack of inversion symmetry is required and different materials have vastly different $\chi^{(2)}$s – hence there still exists opportunity to engineer higher $\chi^{(2)}$s, however, the optimization is still bounded by hard limit of $\chi^{(2)} \sim (\varepsilon-1)/E_0 \sim 10^{-11}-10^{-9} m/V$, once again in the absence of resonance. Note that this limit is achieved in near IR region (III-V semiconductors)[58], but in the visible and UV region there is still room for improvement.

Overall, the simple picture presented here indicates that no dramatic enhancement of off-resonant nonlinearity is possible unless one can "engineer" a potential that is not as smooth as in Fig.2.a Note, that using artificial structures like quantum wells with sharp walls [59] does not yield the desired results as the "effective" potential includes averaging over the electron density (i.e., wavefunction) of electron which effectively leads to the same smooth shape. The only way to avoid is to engineer optically-induced phase transition [60] with a more abrupt change of the potential (Fig.2b) – we shall briefly consider this possibility further down the line, but first we shall return to the key point of this article-role of the interaction time in the nonlinear process.

## Optical nonlinearity in the Feynman diagram picture

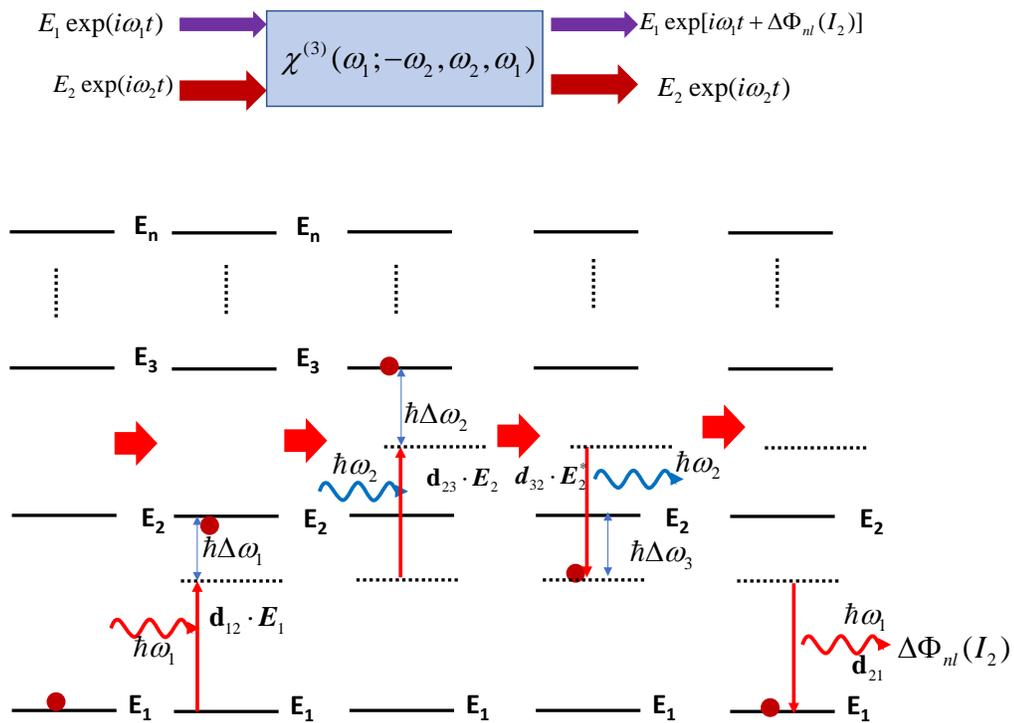

**Figure 3. (a)** Schematics of a typical third order nonlinear optical process -generation of nonlinear polarization at frequency $\omega_1$ and caused by it phase nonlinear phase shift $\Delta\Phi_{nl}$. **(b)** Feynman diagram of this process.

We shall now turn from the classical model for nonlinearity to the quantum one and invoke the language of Feynman diagrams[61]. In this representation n-th order nonlinear process can be represented as sequence on n+1 acts of absorption and emission of photons. For example, linear, or first order process consists of absorption of a photon and subsequent emission of the photon of the same frequency with a phase delay. The second order process, such as sum frequency generation involves consecutive absorption of two photons with frequencies $\omega_1$ and $\omega_2$ followed by emission of as single photon of sum frequency $\omega_3 = \omega_1 + \omega_2$ and so on. (The order of absorption and emission processes can be permuted, and then summation over all possible permutations, or pathways as they are called, must be made, but this important fact is not pertinent to the qualitative discussion here). As an example, consider the cross-phase modulation process, schematically shown in Fig. 2a where nonlinear polarization at frequency $\omega_1$ is proportional to the intensity of light at frequency $\omega_2$ which causes nonlinear phase shift $\Delta\Phi(I_2)$. The Feynman diagram is shown in Fig.2b shows that photons of frequency $\omega_1$ and $\omega_2$ get first absorbed and then emitted with the net result of intensity at one of the frequencies ($I_2 \sim |E_2|^2$) modulating the phase of the other one. The nonlinear polarization can then be written as

$$P^{(3)}(\omega_1;-\omega_2,\omega_2,\omega_1) \sim N\mathbf{d}_{21}\frac{\mathbf{d}_{32}\cdot \boldsymbol{E}_2^*}{\hbar\Delta\omega_3}\frac{\mathbf{d}_{23}\cdot \boldsymbol{E}_2}{\hbar\Delta\omega_2}\frac{\mathbf{d}_{12}\cdot \boldsymbol{E}_1}{\hbar\Delta\omega_1} \tag{1}$$

Now, the probability of m-th step when electron makes transition from energy level $E_m$ to the level $E_{m+1}$ can be evaluated as $\mathbf{d}_{m,m+1}\cdot \boldsymbol{E}_m/\hbar\Delta\omega_m$ where $\mathbf{d}_{m,m+1}$ is the matrix element of a dipole, $E_m$ is electric field of frequency $\omega_m$, and $\hbar\Delta\omega_m = E_{m+1} - E_1 - \sum_{k=1}^{m}\pm\hbar\omega_k$ is the difference between the actual energy acquired by the electron as it resides at an energy level $E_{n+1}$ and the energy received from absorbed (sign +) or transferred to emitted (sign -) photons. One can re-write this expression as $\Omega_m\tau_m$, where $\Omega_m = \mathbf{d}_{m,m+1}\cdot \boldsymbol{E}_m/\hbar$ is the Rabi frequency, describing the rate of interaction between photons and medium and $\tau_m = 1/\Delta\omega_{m+1}$ as the effective interaction time (often called virtual lifetime) which can be understood as such from the uncertainty relation between time and energy. Note that in the off-resonant case, when $\hbar\Delta\omega_m \sim U_0$ one can see that each term $d_{m,m+1}/\hbar\Delta\omega$ in (1) is nothing but inverse of the atomic field $E_0$ as previously defined.

Now, the interaction rate $\Omega_m$ with which the electrons make transition from one level to the next one depends on just one material parameter $\mathbf{d}_{m,m+1}$ whose maximum value is limited, essentially by the size of the bond, i.e. the aforementioned $a$ (of the order of 1-3Å). One can also apply the oscillator sum rule (assuming that most of the strength is associated with a transition at frequency $\omega_{m,m+1} \sim U_0/\hbar$): to obtain $d_{m,m+1} \leq \hbar\sqrt{1/2m_0U_0}$ which results in essentially the same result of $d \leq 2e\times Å$. And this leaves but *one parameter that can be changed within a very wide range, spanning orders of magnitude – the effective interaction time $\tau$ at each step.* One can think of the importance of interaction time using an analogy of a climber on the wall in a gym whose odds of successful climb and descend look better if he or she takes a rest at each step. Let is us now see how it plays out for a typical example of nonlinear phase modulation, and what trade-offs are involved in the process of enhancing nonlinearity via increase in the interaction time.

## Fast and slow nonlinearity

So, it appears that the obvious way to enhance nonlinearity is to increase the interaction time by reducing the detuning $\Delta\omega$, i.e., by taking advantage of the intrinsic resonance of the matter as for instant in atomic vapors[62], or rare-earth dopants[63]. But the resonances remain sharp only for as long as the concentration of the active entities stays low – hence the increase in nonlinearity of each atom or ion gets more than cancelled by the reduction of their number. In a typical solid medium, the resonances are broadened into wide bands, and even the excitonic resonances are broad so one inevitably gets absorption when attempting to approach a resonance. Absorption is the origin of the so-called "slow" nonlinearities [51]when the electron makes a "real" optically induced transition between the bands or within the same band. Once the electron is excited several different processes can take place such as absorption saturation[64], exciton screening[15], rise of electron temperature with subsequent change of effective mass [65], or simply the rise of the lattice temperature[66]. Each of these processes is accompanied by change of the index of refraction and is characterized by its own characteristic time $\tau$, which can be recombination time, intraband scattering time, thermal diffusion time, and so on. The range of characteristic times is very wide – from tens of femtoseconds in intraband processes to milliseconds for thermal diffusion times with recombination rates being anywhere from picoseconds to tens of microseconds. The energy density stored in the medium can be estimated as $\Delta u_{st} \sim \alpha I \tau$, where $\alpha$ is absorption coefficient and $I$ is power density of propagating light (it can be the power of the signal for the self-phase modulation case, or the power of control light $I_2$ for the cross-phase modulation case of Fig.3). Then, the order of magnitude of the optically induced index change is $\Delta n_{slow} \sim \alpha n I \tau / 2 u_0$, and the maximum phase shift is determined by the propagation length equal to $1/\alpha$, i.e.

$$\Delta \Phi_{slow} \sim k_0 n I \tau / 2 u_0 \qquad (2)$$

Let us now find the phase shift imposed by the "fast" nonlinearity, i.e. far away from resonance,

$$\Delta \Phi_{fast} = \frac{1}{2n}\chi^{(3)}E^2 k_0 L \sim \frac{n^2-1}{2n^2}\frac{e^2 a^2}{U_0^2}\eta_0 I k_0 L, \qquad (3)$$

where $\eta_0 = 377\Omega$ is vacuum impedance and $L$ is the propagation length, which, in the absence of absorption can be arbitrarily long. The "fast" nonlinearity occurs on a time scale of $1/\Delta\omega$ i.e., far from resonance on sub-femtosecond scale, and, for all practical reasons can be considered nearly "instant". We can now establish a connection between "fast" and "slow" phase shifts. Since linear susceptibility away from the resonance can be estimated as $\chi^{(1)} = n^2 - 1 \sim Ne^2 a^2 / \varepsilon_0 U_0$ one can re-write (2) as

$$\Delta \Phi_{slow} = \frac{1}{2}\frac{n}{n^2-1}\frac{e^2 a^2 \tau}{\varepsilon_0 U_0^2}k_0 I \qquad (4)$$

The ratio of nonlinear phase shift achieved by fast and slow mechanisms is then

$$\frac{\Delta\Phi_{fast}}{\Delta\Phi_{slow}} = \frac{(n^2-1)^2}{n^3}\frac{\varepsilon_0\eta_0 L}{\tau} = \frac{(n^2-1)^2}{n^4}\frac{nL}{c\tau} \sim \frac{\tau_{tr}}{\tau}, \quad (5)$$

where $\tau_{tr} = Ln/c$ is the *transit(propagation) time*. Therefore, no matter whether the nonlinearity is instant or "slow" the nonlinear index change can always be written as

$$\Delta\Phi = \frac{1}{2}\frac{n}{n^2-1}\frac{\tau_{eff}}{\varepsilon_0 E_0^2} k_0 I \quad (6)$$

where *the effective interaction time $\tau_{eff}$ is either a propagation time in case of fast ("instant") nonlinearity or some actual intrinsic relaxation time in case of "slow" nonlinearity.* The switching intensity that provides the required $\pi/2$-shift is then

$$I_{\pi/2} \sim 2n\varepsilon_0 E_0^2 \lambda_0 / \tau_{eff} \quad (7)$$

The key feature of (6) and (7) is that *$\tau_{eff}$ is the only parameter that can be changed by orders of magnitude and one can enhance nonlinear shift (reduce $I_{\pi/2}$), but with inevitable tradeoffs that are actually different for slow and fast nonlinearities.* For "slow" nonlinearity the actual *speed* (or bandwidth) of the nonlinear device gets reduced as $1/\tau_{eff}$, while for "instant" nonlinearity increase in transit time only increases the *latency*, which in many applications is less deleterious than decrease in speed. Another important difference between "slow" and "fast" nonlinearities is that the energy of light is dissipated in "slow" absorption-based nonlinearity but is preserved in case of fast nonlinearity and can in principle be reused. Overall, one can write for the product of switching intensity and delay time (which can be thought of as inverse of the gain-bandwidth product used in electronics) as

$$I_{\pi/2}\tau_{eff} \sim n\varepsilon_0 E_0^2 \lambda_0 \sim 10^1 - 10^3 \, GW \cdot ps/cm^2 \quad (8)$$

with lower numbers corresponding to mid-IR and higher numbers to visible-UV. Here we need to emphasize that while for "slow" nonlinearity (8) can be interpreted as switching energy (per unit area), no such connection can be made for fast nonlinearity in traveling wave geometry where the switching energy would be simply determined by the bit length $U_{sw} \sim I_{\pi/2}T_{bit}$, i.e., reduced considerably if the bit duration much shorter than transit time (as is the case in nonlinear fiber optics)

## Enhancement: intrinsic or extrinsic?

When it comes to enhancement of nonlinearity via increase of interaction time, one may distinguish between intrinsic and extrinsic methods. For "fast" nonlinearity the intrinsic enhancement consists of using the resonances inside the material. As we have already discussed, approaching resonance, and still staying away from absorption requires very narrow resonances, attainable in such media as atomic vapors, but not in condensed matter. Even using excitonic resonances[16, 67] is not going to enhance the nonlinear phase shift by a lot. This is due to the fact that excitons typically contain only the states near the center of the Brillouin zone, i.e., only a small fraction of all states, hence while the excitonic effects on imaginary part of permittivity (absorption) are indeed significant, the real part of permittivity is not strongly affected. One can take this argument even further and mention that other peculiarities of band structures, such as Dirac[26] and Weyl[30] points hardly affect refraction index and real part of nonlinear susceptibility simply because they represent but a tiny fraction of all the states contributing to the $n$ and real $\chi^{(3)}$.

At the same time, in slow nonlinearity engineering the characteristic relaxation time $\tau$ makes perfect sense. When the required switching time is $t_{sw}$, it makes sense to keep $\tau$ equal to a fraction of it, say $\tau \sim 0.2 - 0.5 t_{sw}$ and not to any shorter, as shortening will only reduce the nonlinearity. With today's requirements of $t_{sw} \sim 10 ps$ (for 100GBps communications) and prospectively going to $t_{sw} \sim 1 ps$, most of the nonlinearities based on interband transitions are too slow, although one may consider low temperature grown materials that are defect rich in which recombination times are reduced to within a picoseconds range. On the other hand, the intraband relaxation in most materials occurs on the scale of 100's of femtoseconds. For example, hot carriers in transparent conductive oxides (TCO's.[31]) have been shown to thermalize in sub picosecond times. Previopusly the same relaxation times were observed in III-V semiconductors using spectral hole burning measurements[68]. That is part of the reason why TCO's have become subjects of such a strong interest lately. One can also easily engineer the relaxation times for intersubband transitions in quantum wells [69]

One should note that in terms of their nonlinear indices "slow" nonlinearities are always stronger than "fast" ones by ratio of relaxation time $\tau$ to the coherence (or dephasing time) $T_{coh}$ [51], which can be orders of magnitude, but, as we have emphasized throughout this paper, it is not nonlinear index but the nonlinear phase shift that matters. Therefore, according to (6) one can achieve comparable results in "slow medium" with relaxation time of 1ps and couple of hundreds micron long "fast" medium with 1ps propagation time. The advantage of slow medium would be of course the smaller size since absorption length is only a few microns. But the disadvantage would be in heat dissipation. Since the threshold of thermal damage is much lower than optical breakdown, it is questionable whether such schemes as based on TCO can operate at high duty cycle required for bandwidth efficient communications and processing[70]. (Typically, the experiments with sub-picosecond all-optcal switching[71, 72] are performed using sub-picosecond pulses from the mode-locked lasers operating at less than 100MHz setting maximum data throughput at 100Mbit/s, i.e., a far cry from Tbit/s as one would naively believe).

A different (and complementary) approach to interaction time enhancement is doing it externally. The most straightforward method is using the so-called "slow light effect" [21]in highly dispersive medium, such as the TCO near ENZ point[73]. When dispersion is considered, the transit time is $\tau_{tr} = L n_g / c$, where $n_g = c dk/d\omega$, is a group index, which can be of the order of 1 to 10, hence one can attain the same phase shift in a shorter length. Similar effect is achievable with photonic crystals [74](which of course requires additional fabrication steps) and plasmonic waveguides[75] (plagued by loss). By far the most widely used way of enhancing the interaction time is by using one or another type of a resonant photonic structure. In a simple Fabry-Perot or ring resonator the effective interaction time is enhanced by a finesse $F$, which is roughly a number of round trips photon makes inside the cavity before leaking out (or being absorbed). Alas, this approach has a major shortcoming as interaction time increase is accompanied by the reduction in the bandwidth (and not latency! ) But at the same time there is an additional benefit of increasing the electric field of the pump wave as the energy gets "compressed" inside the resonator.

What is important to mention is that recent years have seen multiple sophisticated resonant schemes designed to achieve "giant enhanced nonlinearity". One can mention Fano resonances[76], electromagnetically induce transparency (EIT) [22], all kinds of nanoantennas and metasurfaces[77], photonic bound states in continuum[78], anapoles[79], topological modes[32], and this list can be continued ad infinitum. All said and done; no matter which scheme is used the effective interaction time for nonlinearity is simply $\tau_{eff} = Q/\omega$, and $Q$ factor in most of the sophisticated schemes does not exceed Q of a simple micro resonator[80] or a ring resonator[81], and, at any rate it is mostly determined by the intrinsic absorption. It seems that with $Q \sim 10^3$ one may achieve effective interaction times on the scale of few *ps* – any further increase would affect the bandwidth. It can also be shown that in many other sophisticated schemes of enhancing nonlinearity such as enhancement near exceptional points[82] one can also relate that increase to the interaction time.

Before turning to conclusions, it is instructive to emphasize the importance of interaction time using the example of "cascaded" nonlinearity, i.e., using the second order processes sequentially, rather than third order process[83]. The most widely known example is generating of third harmonic by first frequency doubling the fundamental frequency $\omega$ and then mixing the second harmonic $2\omega$ with remaining fundamental to obtain $3\omega$ [84]. This cascaded process, known to be far more efficient than directly generating the third harmonic, is the operating principle of frequency tripling in all commercial laser systems. But cascading can also emulate the entire spectrum of third order nonlinear process, such as nonlinear phase shift[85], four wave mixing[86] and others. As shown in [83] the in cascaded nonlinearity the intensity required for 90 degrees phase shift is $I_{\pi/2}^{(casc)} \sim I_{\pi/2}/\omega\tau_{tr}$, i.e. effective nonlinearity gets enhanced by the propagation factor $Q = \omega\tau_{tr}$. Obviously, if one uses a resonant cavity, then the enhancement is just the cavity Q. As explained in [83] it is achieved by coupling one of the short-lived "virtual" excitations in Fig.3 to the photon at intermediate frequency and thus extending its lifetime.

## Conclusions

In this (originally intended to be short) discourse I have attempted to emphasize the role of what can be broadly defined as an interaction time in the nonlinear optics. Since photons do not interact with each other directly but only via excitation of matter (real or virtual) and the strengths of this interaction, depends only on value of the Hamiltonian determining the interaction and the lifetime of the excitation (real or virtual) referred here as interaction time. The strongest photon-matter interaction is the dipole transition, and its maximum value is limited by what is typically a bond length, at most a couple of Angstroms.

That leaves only the interaction time as a flexible parameter capable of enhancing effective nonlinearity. This enhancement can be achieved intrinsically, typically by operating in the absorption region and exciting real carriers with characteristics lifetimes that can vary by orders of magnitude. The inevitable tradeoff then reduces the switching speed, but even though the nonlinearity is "slow" it can be sufficiently fast for a given application. TCO materials capable of operating at less 1ps is a prime example of such nonlinearity.

It is far less feasible to intrinsically enhance interaction time for the so-called "ultra-fast" nonlinearities operating away from absorption, i.e., with only virtual carriers being excited, mostly because in practical nonlinear medium the resonances are very broad as individual energy levels broaden into the bands.

The broadening is inevitable, because the photon-matter interactions are weak, and many atoms(ions) are required to achieve appreciable effect without absorption. *This last fact also causes me to be a bit skeptical about the prospects of widely lauded low-dimensional materials as well as topological materials – a few states in a single atomic layer or in the vicinity of a special point in Brillouin zone may have extraordinary nonlinear properties, but since there are only a few of them, the overall yield y of nonlinear process (phase shift) remains pitifully low.*

With both slow and especially with ultrafast nonlinearities, it is the extrinsic enhancement of interaction time that is bound to yield the best results. The enhancement can be achieved in both propagating (travelling wave) and localized geometry. The most obvious way to achieve enhancement is by simply using long propagation lengths as in fibers, where, despite all the hype surrounding other methods, the best results for all-optical switching, modulation, and frequency conversion have been achieved to date. To reduce the actual physical length while keeping interaction time long one may revert to various photonic structures low group velocity ("slow light"), a good example being photonic crystal waveguides[87]. Using travelling wave geometry allows one to enhance the nonlinearity while keeping bandwidth reasonably wide and only increasing the latency, although in the end increasing delay also reduces the bandwidth due to group velocity dispersion[88]

Alternatively, various photonic resonant structures may be employed for enhancement. Despite their proliferation over the past decade, it is my opinion that a simple micro-resonator remains the best option for practical enhancement. The degree of the enhancement in localized resonant structures is ultimately limited by the reduction of the bandwidth, hence the ultra-high Q-factors might not be required.

I have also shown that cascaded second order nonlinearity may achieve the same goals as third order nonlinearity with significant enhancement due to longer interaction time. But implementation of cascading requires phase-matching and thus significantly complicates the effort. In the end one always should be able to judiciously choose which ("slow" or "fast") nonlinearity to use and how to enhance it based on demand of each concrete application, for example whether the emphasis should be placed on speed, size, or low loss.

As far as new entrants into the rather stale menu of practical nonlinear materials are concerned, there is plenty of room for development of materials that have very desirable properties of high damage threshold, low insertion loss, low cost, ability to be integrated on Si platform, and many others. But it is my view that fundamental laws of physics would not allow enhancements of efficiencies (or reducing power requirements) for a given wavelength and required speed by orders of magnitude, notwithstanding all the anticipated future claims of "giant nonlinearities" that will sure continue to appear with admirable regularity. Not to sound too dire, perhaps one way to look at the enhancement of strength rather than interaction time is to investigate exploiting phase transitions in which case the effective potential (Fig.1b) becomes anharmonic well below optical damage threshold, but this transition needs to be of electronic rather than ionic nature for operating in the near IR to visible region. Potential candidates may be optically induced Mott-Andersen localization transition[89] or other collective effects in solids, such as Wigner crystal formation[90], assuming (big if) they can be observed above cryogenic temperatures.

To summarize, the goal of this write-up has been to provide a view from a less familiar angle at the state of developments in nonlinear optics and point out the key factors that limit the efficiency of nonlinear optical phenomena and prevent nonlinear optics from entering new applications, especially in all optical data processing. It is my hope that this goal has been to a certain degree realized, and that the universal language of "interaction time engineering" introduced here will be used to enhance and customize nonlinear phenomena in the future. On whether all optical switching and processing based on nonlinear optics will ever materialize I will withhold my opinion as many unexpected discoveries may occur and new demands may arise in not-so-distant future.

Acknowledgement

The author appreciates valuable support by DARPA and he is as always greatly obliged to the unrelenting encouragement and assistance by his co-workers Prof. P. Noir and Dr. S. Artois without whom this feat could hardly become possible.